\documentclass{article}

\usepackage{microtype}
\usepackage{graphicx}
\usepackage{subfigure}
\usepackage{booktabs} % for professional tables

\usepackage{hyperref}

 \usepackage[accepted]{synsml2023}

\usepackage{amsmath}
\usepackage{amssymb}
\usepackage{mathtools}
\usepackage{amsthm}

\usepackage{commath}
\usepackage[utf8]{inputenc}
\usepackage[T1]{fontenc}
\usepackage{mathptmx}
\usepackage{etoolbox}

\usepackage{xcolor}
\usepackage{hyperref}

\usepackage[capitalize,noabbrev]{cleveref}

\theoremstyle{plain}

\theoremstyle{definition}

\theoremstyle{remark}

\usepackage[textsize=tiny]{todonotes}

\synsmltitlerunning{Submission and Formatting Instructions for Syns \& ML at ICML 2023}

\begin{document}

\twocolumn[
\synsmltitle{Multi-Fidelity Data Assimilation For Physics Inspired Machine Learning In Uncertainty Quantification Of Fluid Turbulence Simulations}

% unless you've provided the [accepted] option to the synsml2023


\synsmlsetsymbol{equal}{*}

\begin{synsmlauthorlist}

\synsmlauthor{Minghan Chu}{equal,yyy}
\synsmlauthor{Weicheng Qian}{equal,comp}

\synsmlauthor{Department of Mechanical and Materials Engineering}{yyy}
\synsmlauthor{Department of Computer Science}{comp}

\synsmlauthor{Queen's University}{yyy}
\synsmlauthor{University of Saskatchewwan}{comp}

\synsmlauthor{\texttt{17MC93@queensu.ca}}{yyy}
\synsmlauthor{\texttt{weicheng.qian@usask.ca}}{comp}

%\synsmlauthor{Firstname1 Lastname1}{equal,yyy}
%\synsmlauthor{Firstname2 Lastname2}{equal,yyy,comp}
%\synsmlauthor{Firstname3 Lastname3}{comp}
%\synsmlauthor{Firstname4 Lastname4}{sch}
%\synsmlauthor{Firstname5 Lastname5}{yyy}
%\synsmlauthor{Firstname6 Lastname6}{sch,yyy,comp}
%\synsmlauthor{Firstname7 Lastname7}{comp}
%%\synsmlauthor{}{sch}
%\synsmlauthor{Firstname8 Lastname8}{sch}
%\synsmlauthor{Firstname8 Lastname8}{yyy,comp}
%%\synsmlauthor{}{sch}
%%\synsmlauthor{}{sch}
\end{synsmlauthorlist}

%\synsmlaffiliation{yyy}{Department of XXX, University of YYY, Location, Country}
%\synsmlaffiliation{comp}{Company Name, Location, Country}
%\synsmlaffiliation{sch}{School of ZZZ, Institute of WWW, Location, Country}

%\synsmlcorrespondingauthor{Firstname1 Lastname1}{first1.last1@xxx.edu}
%\synsmlcorrespondingauthor{Firstname2 Lastname2}{first2.last2@www.uk}

%% You may provide any keywords that you
%% find helpful for describing your paper; these are used to populate
%% the "keywords" metadata in the PDF but will not be shown in the document
%\synsmlkeywords{Machine Learning}

\vskip 0.3in
]


% The \synsmlEqualContribution command is standard text for equal contribution.

%\printAffiliationsAndNotice{\synsmlEqualContribution} % otherwise use the standard text.

\begin{abstract}
Reliable prediction of turbulent flows is an important necessity across different fields of science and engineering. In Computational Fluid Dynamics (CFD) simulations, the most common type of models are eddy viscosity models that are computationally inexpensive but introduce a high degree of epistemic error. The Eigenspace Perturbation Method (EPM) attempts to quantify this predictive uncertainty via physics based perturbation in the spectral representation of the predictions. While the EPM outlines how to perturb, it does not address how much or even where to perturb. We address this need by introducing machine learning models to predict the details of the perturbation, thus creating a physics inspired machine learning (ML) based perturbation framework. In our choice of ML models, we focus on incorporating physics based inductive biases while retaining computational economy. Specifically we use a Convolutional Neural Network (CNN) to learn the correction between turbulence model predictions and true results. This physics inspired machine learning based perturbation approach is able to modulate the intensity and location of perturbations and leads to improved estimates of turbulence model errors. 
\end{abstract}

\section{Introduction}
\label{submission}
Turbulent fluid flows are common across different fields of science and engineering. The most commonly used models are Eddy Viscosity Models (EVMs) that are computationally inexpensive but introduce a high degree of epistemic error or model-form uncertainty, due to simplifications like the Boussinesq Turbulent Viscosity Hypothesis (TVH). The quantification of these model-form uncertainty \cite{duraisamy2017status} ensures reliable designs and analysis. The Eigenspace Perturbation Method (EPM) \cite{emory2013modeling,iaccarino2017eigenspace} is a physics based framework to quantify model-form uncertainty introduced in EVMs. This method has shown reliable uncertainty quantification with low computational cost, and has been applied to a variety of engineering problems including virtual certification of aircraft designs \cite{mukhopadhaya2020multi, nigam2021toolset}, design of urban structures\cite{gorle2019epistemic}, aerospace design and analysis\cite{mishra2019uncertainty, mishra2017rans, mishra2019estimating, mishra2017uncertainty}, application to design under uncertainty (DUU) \cite{demir2023robust, cook2019optimization, mishra2020design, righi2023uncertainties}, etc . 

The EPM only outlines how to perturb, while it does not address how much to perturb. As an illustration the EPM uses the maximal physics permissible perturbation all over the flow domain. The strength of perturbation should reflect the degree of discrepancy between EVMs and the truth, as the discrepancy actually varies spatially in the flow domain. We address this need by developing a function to predict the degree of perturbation at different locations. While there aren't physics based precepts to completely determine this function to predict the spatial variation of perturbations, it can be estimated reliably from data. Machine Learning based approaches are becoming more used in fluid mechanics applications \cite{duraisamy2019turbulence, chung2021data, brunton2020machine}. Some prior investigators have tried to use ML to improve turbulence model UQ \cite{xiao2016quantifying,wu2018physics,heyse2021estimating,heyse2021data,zeng2022adaptive}. These studies have focused on more complex models that necessitate large incorporation of labeled data and limit generalizability. These studies used polynomial regression \cite{chu2022model}, Probabilistic Graphical Models (PGMs), Random Forests, etc. These do not reflect the non-locality of turbulence physics where the evolution of a turbulent flow at a location is affected by far off points in the flow domain too. 

We utilize a new solution which tries to compensate for the absent non-local physics in turbulence models via the inductive biases of Convolutional Neural Networks (CNNs). We use this to formulate a function that predicts the degree of perturbation and improving the EPM. 

\section{Methodology}

\begin{figure*}[t]
    \centering
    \includegraphics[width=\linewidth]{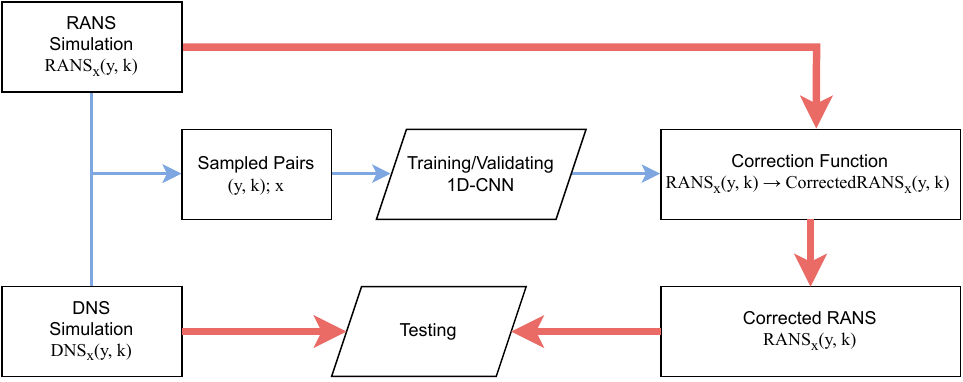}
    \caption{Data-flow Diagram for Experiments. Blue path is the training/validating path, and the red path is the validation path. Blue path represents small amount (about 20\%) of data through flow. Red path represents large amount (about 80\%) of data through flow.} %Thickness of the path indicates the amount of data through the flow. }
    \label{fig:data-flow.pdf}
\end{figure*}

The turbulence kinetic energy $k$ plays an important role in the estimates of the model-form uncertainty introduced in RANS models using the Eigenspace Perturbation Method \cite{emory2013modeling,iaccarino2017eigenspace}. We employed a one-dimensional convolutional neural network (1D-CNN) to learn the function to correct RANS prediction towards DNS prediction, namely, correction function. Our approach enables the use of learned correction function for the computationally cheaper RANS prediction to replace the rather expensive high-fidelity DNS prediction. 

For both the RANS and DNS simulation, we can summarize their results as the function of the perturbed turbulence kinetic energy $k^{*} = f(x,y)$, where $x$ and $y$ are coordinates in a two-dimensional computational domain, and $f$ is the mapping from every coordinate $(x, y)$ to $k^{*}$, embedded in triples $(x, y, k^{*})$ from simulation results.

%For sake of simplicity, we will use k to represent k^{*} throughout the following sections. 
Without assuming a specific form, the correction function for RANS is a mapping between two triples $\zeta: (x, y, k^{\mathrm{RANS}}) \rightarrow (x, y, k^{\mathrm{DNS}})$. Consider the model error for RANS and DNS in terms of kinetic energy, we have
\begin{equation}
   p^{\text {RANS }}\left(K_g \mid x, y\right)=p\left(k_g=k^{\text {RANS }} \mid x, y\right),
\end{equation}

\begin{equation}
    p^{\text {DNS }}\left(K_g \mid x, y\right)=p\left(k_g=k^{\text {DNS }} \mid x, y\right),
\end{equation}

where $K_g$ is the unknown ground truth of kinetic energy at $(x, y)$.
Turbulence kinetic energy that results from DNS simulation, i.e., $p^{\mathrm{RANS}}$, can be estimated using RANS-based turbulence kinetic energy $p^{\mathrm{DNS}}$ and its correction function $g$ as

\begin{equation}
p^{\mathrm{DNS}}\left(K_g \mid x, y\right)=g\left(k^{\mathrm{RANS}}, x, y\right) p\left(k^{\mathrm{RANS}} \mid x, y\right).
\end{equation}

Because $k^{\mathrm{DNS}} = f^{\mathrm{DNS}}(x, y)$ and $k^{\mathrm{RANS}} = f^{\mathrm{RANS}}(x, y)$,
at each $x$, we have $k_x^{\mathrm{DNS}} = f_x^{\mathrm{DNS}}(y)$ and $k_x^{\mathrm{RANS}} = f_x^{\mathrm{RANS}}(y)$, assuming both $f_x^{\mathrm{RANS}}$ and $f_x^{\mathrm{DNS}}$ are continuous, that is, $\forall \epsilon > 0,\allowbreak\, \exists \delta > 0,\allowbreak\, s.t.\, \forall \abs{d} < \delta, \allowbreak \, \abs{f_x(y + d) - f_x(y)} < \epsilon$. We can approximate $g(k^{\mathrm{RANS}}, x, y)$ with $\hat{g}(\mathbf{k}_{x,y,\delta}^{\mathrm{RANS}})$, where $\mathbf{k}_{x,y,\delta}^{\mathrm{RANS}} = [k_{x, y_0}^{\mathrm{RANS}}, k_{x, y_1}^{\mathrm{RANS}}, \cdots]^\top$ and $y_0, y_1, \dots \in [y - \delta, y + \delta]$. In other words, we can learn $\hat{g}$ with paired $(\mathbf{k}_{x,y,\delta}^{\mathrm{RANS}}, \mathbf{k}_{x,y,\delta}^{\mathrm{DNS}})$.

Our one-dimensional convolutional neural network (1D-CNN) learns the correction function $\hat{g}$ from paired RANS and DNS simulation estimated turbulence kinetic energy $(\mathbf{k}_{x,y,\delta}^{\mathrm{RANS}}, \mathbf{k}_{x,y,\delta}^{\mathrm{DNS}})$. Because our approximated correction function $\hat{g}$ only depends on the neighbor of $k^{\mathrm{RANS}}$, and coordinates $(x, y)$ are only used to group neighbors of $k^{\mathrm{RANS}}$, we grouped simulation data by $x$ and transformed $(y, k)$ at $x$ into $\mathbf{k}_{x,y,\delta}^{\mathrm{RANS}}$ via a rolling window parameterized by window size. Our 1D-CNN has four-layers and in total 86 parameters: a single model for all zones at any $x$ to correct RANS towards DNS.

% \begin{figure*}[h!]
%     \centering
%     \includegraphics[width=\linewidth]{fig/Schematics_SD7003_hills.pdf}
%     \caption{(a) flow passing over an SD7003 airfoil and (b) flow over a two-dimensional periodically arranged hills.} %Thickness of the path indicates the amount of data through the flow. }
%     \label{fig:Schematics_SD7003_hills.pdf}
% \end{figure*}

\section{Experiments Setup and Data Sources}
We proposed a lightweight 1D-CNN approach to approximate the correction function that corrects RANS simulation towards DNS simulation. We experimented our lightweight CNN-based approach to approximate the correction function for RANS on two datasets: the in-house RANS/DNS \cite{zhang2021turbulent,chu2022model} dataset and the public RANS/DNS dataset \cite{voet2021hybrid}. The in-house RANS/DNS dataset \cite{chu2022model,zhang2021turbulent} was obtained by considering the flow around an SD7003 airfoil. The public RANS/DNS dataset \cite{voet2021hybrid} was generated from the two-dimensional channel flow over periodically arranged hills. Both flow cases experience adverse pressure gradient, which causes the complex flow features of separation and reattachment. It is challenging both to understand and to model.

From \cref{fig:data-flow.pdf}, we split $x$-coordinate grouped pairs of $(\mathbf{k}_{x,y,\delta}^{\mathrm{RANS}}, \mathbf{k}_{x,y,\delta}^{\mathrm{DNS}})$ into training set and validating set by their group key $x$. For both the in-house and the public dataset, we choose $x$ at only three positions on the geometry from the beginning, the middle, and the end of all paired $x$ values. For the in-house dataset based on the SD7003 airfoil geometry, $x/c = 0.4, 0.56, 0.58$; for the public dataset based on the two-dimensional periodically arranged hills, $x/c = 0, 0.046, 0.116, 0.128$, where $c$ is a reference length used for normalization. For each dataset, we use a 80\%--20\% split as training--testing dataset.

For both datasets, we validated our trained 1D-CNN by comparing the L1 loss of RANS, denoted $L^1_c(\texttt{rans}) = \abs{ CF^{\mathrm{RANS}}_{k} - CF^{\mathrm{DNS}}_{k} }$, with the L1 loss of 1D-CNN corrected RANS, denoted $L^1_c(\texttt{pred}) = \abs{ CF^{\mathrm{CNN}}_{k} - CF^{\mathrm{DNS}}_{k} }$.

%Furthermore, we examined the generalizability of our lightweight CNN-based correction function trained on the public RANS/DNS dataset \cite{voet2021hybrid} by applying the correction function for other cases. 

\section{Results and discussion}
\begin{figure*}[!htb]
%outer boundary layer     
    \centering
    \includegraphics[width=\linewidth]{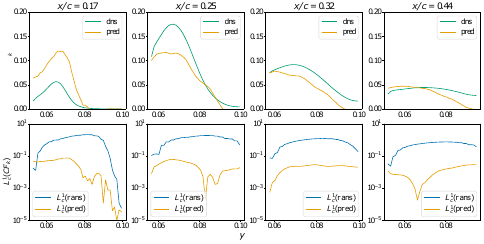}
    \caption{Results for Selig-Donovan 7003 airfoil. First row: CNN corrected DNS (\texttt{pred}) compared with ground truth (\texttt{dns}). Second row: Validation of 1D-CNN by comparing L1 loss between $L^1_c(\texttt{rans})$ and $L^1_c(\texttt{pred})$.}
    \label{fig:cnn-corrected-rans-zhang.pdf}
\end{figure*}

\begin{figure*}[!htb]
%outer boundary layer     
    \centering
    \includegraphics[width=\linewidth]{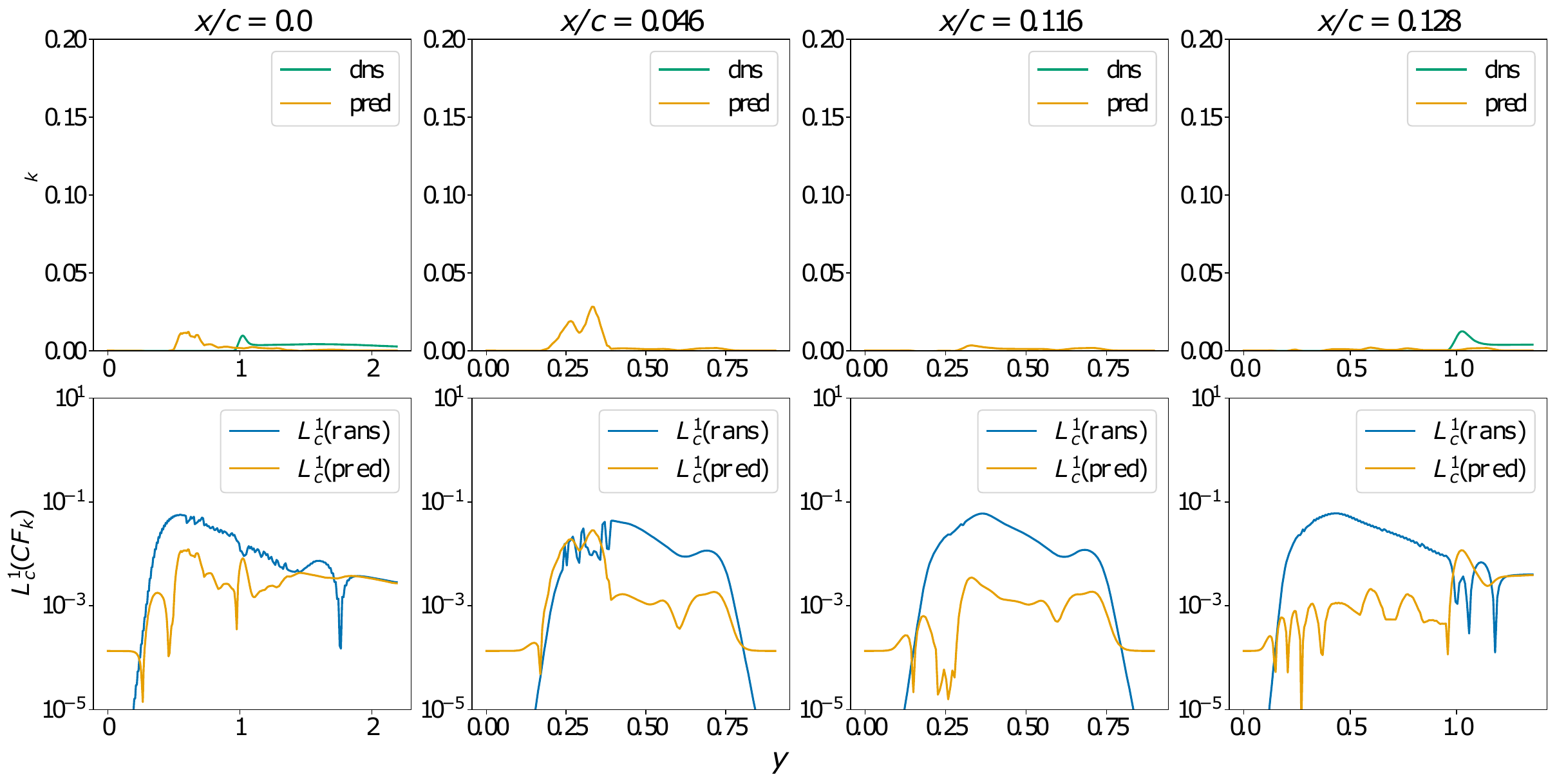}
    \caption{Results for two-dimensional periodically arranged hills. First row: CNN-based prediction (\texttt{pred}) compared with ground truth (\texttt{dns}). Second row: Validation of 1D-CNN by comparing L1 loss between $L^1_c(\texttt{rans})$ and $L^1_c(\texttt{pred})$.}
    \label{fig:cnn-corrected-rans-voet.pdf}
\end{figure*}

Our CNN-based correction function is validated at all paired $x$ locations. \Cref{fig:cnn-corrected-rans-zhang.pdf,fig:cnn-corrected-rans-voet.pdf} show the results for SD7003 and 2D periodically arranged hills, respectively, at four $x$ locations within the region of separated flow. From \cref{fig:cnn-corrected-rans-zhang.pdf}, the series of CNN predicted DNS profiles in the first row are then smoothed with the moving average (the average of points in a sliding window) with a window size of six consecutive estimations. Our CNN-based prediction for the $k$ profile resemble the ground truth DNS despite being trained with only a few pairs of RANS and DNS results. 

For both datasets, the CNN-based correction function is trained on paired RANS-DNS simulated turbulence kinetic energy using less than $20\%$ positions along $x$-axis, while the CNN-based function is still effective for the remaining $80\%$ positions. In other words, our CNN-based correction can be used to predict RANS-DNS simulated turbulence kinetic energy at any $x$ coordinate with only a fraction of the whole $x$ coordinates. Furthermore, our lightweight CNN model uses the $y$ coordinates for grouping RANS-simulated turbulence kinetic energy within a neighbor. The results of our CNN-based correction function suggests that RANS results might be improved by leveraging information embedded in the positions within a close neighbor, which is independent of the absolute coordinates $(x, y)$. The lightweight CNN-based correction function trained on one case can still help smooth and reduce the error of RANS-based results for other cases.

For both the in-house and the public flow cases, our CNN-based prediction for $k$ lies closer to the DNS data at any $x$ location, i.e., the discrepancy in general reduces as the flow proceeds further downstream. The RANS results deviate from the DNS data in both flow scenarios and our CNN-based correction function can significantly reduce the L1 error of RANS- from DNS-simulations, as shown in the second row of \cref{fig:cnn-corrected-rans-zhang.pdf,fig:cnn-corrected-rans-voet.pdf}, i.e., a $L^1_c(\texttt{pred})$ drop of two orders compared to $L^1_c(\texttt{rans})$. 

From \Cref{fig:cnn-corrected-rans-zhang.pdf,fig:cnn-corrected-rans-voet.pdf}, it is interesting to note that the CNN-based prediction for $k$ tends to approach closer to the DNS profile as the flow proceeds further downstream. This indicates that the CNN-based correction function tends to become more trustworthy within the region of fully turbulent flow where flow features are less complex than that within region of separated flow where rather complex flow features evolve.


\section{Conclusion}
\label{sec:Conclusion}

To the best of our knowledge, we are among the first to examine the projection from RANS to DNS using the CNN approach. Our experiment results suggest that the CNN-based correction function corrects the RANS predictions for perturbed turbulence kinetic energy towards both the in-house and the public DNS data. A projection that can approximate the in-house DNS data reasonably well from RANS might exist independent of $x$. Our methodology can be easily extended to analyze flows over different types of airfoils.

Our findings are subject to following limitations: our CNN model was trained with only two datasets. Future work may include validation with more datasets on different flow cases, e.g., different types of airfoils.  In addition, our CNN-based correction function will be integrated with the eigenspace perturbation framework to result in accurate perturbations and hence improved estimation of RANS UQ.

\section{Impact statement}
Separation, reattachment, and separation bubbles are so frequently encountered in real life engineering applications. However, those flow features are complex to both understand and model. LES and DNS give high-fidelity results, but the calculations are too expensive for practical engineering applications. Therefore, knowing the uncertainty associated with the low-cost RANS simulations are of great interest as it builds confidence when using RANS. An effective physics-based eigenspace perturbation method only presupposes the worst case scenario where maximum strength (turbulence kinetic energy) of perturbation is assumed. Thus it gives overly conservative uncertainty estimates. Our physics inspired machine learning based perturbation approach is able to spatially modulate the strength and location of the physics based perturbations and leads to improved estimates of turbulence model errors.

\nocite{langley00}

\bibliography{example_paper}

\providecommand{\noopsort}[1]{}\providecommand{\singleletter}[1]{#1}%
\begin{thebibliography}{25}
\providecommand{\natexlab}[1]{#1}
\providecommand{\url}[1]{\texttt{#1}}
\expandafter\ifx\csname urlstyle\endcsname\relax
  \providecommand{\doi}[1]{doi: #1}\else
  \providecommand{\doi}{doi: \begingroup \urlstyle{rm}\Url}\fi

\bibitem[Brunton et~al.(2020)Brunton, Noack, and
  Koumoutsakos]{brunton2020machine}
Brunton, S.~L., Noack, B.~R., and Koumoutsakos, P.
\newblock Machine learning for fluid mechanics.
\newblock \emph{Annual Review of Fluid Mechanics}, 52:\penalty0 477--508, 2020.

\bibitem[Chu et~al.(2022)Chu, Wu, and Rival]{chu2022model}
Chu, M., Wu, X., and Rival, D.~E.
\newblock Model-form uncertainty quantification of reynolds-averaged
  navier--stokes modeling of flows over a sd7003 airfoil.
\newblock \emph{Physics of Fluids}, 34\penalty0 (11):\penalty0 117105, 2022.

\bibitem[Chung et~al.(2021)Chung, Mishra, Perakis, and Ihme]{chung2021data}
Chung, W.~T., Mishra, A.~A., Perakis, N., and Ihme, M.
\newblock Data-assisted combustion simulations with dynamic submodel assignment
  using random forests.
\newblock \emph{Combustion and Flame}, 227:\penalty0 172--185, 2021.

\bibitem[Cook et~al.(2019)Cook, Mishra, Jarrett, Willcox, and
  Iaccarino]{cook2019optimization}
Cook, L.~W., Mishra, A., Jarrett, J., Willcox, K., and Iaccarino, G.
\newblock Optimization under turbulence model uncertainty for aerospace design.
\newblock \emph{Physics of Fluids}, 31\penalty0 (10):\penalty0 105111, 2019.

\bibitem[Demir et~al.(2023)Demir, Gorguluarslan, and Aradag]{demir2023robust}
Demir, G., Gorguluarslan, R.~M., and Aradag, S.
\newblock Robust shape optimization under model uncertainty of an aircraft wing
  using proper orthogonal decomposition and inductive design exploration
  method.
\newblock \emph{Structural and Multidisciplinary Optimization}, 66\penalty0
  (4):\penalty0 93, 2023.

\bibitem[Duraisamy et~al.(2017)Duraisamy, Spalart, and
  Rumsey]{duraisamy2017status}
Duraisamy, K., Spalart, P.~R., and Rumsey, C.~L.
\newblock Status, emerging ideas and future directions of turbulence modeling
  research in aeronautics.
\newblock Technical report, 2017.

\bibitem[Duraisamy et~al.(2019)Duraisamy, Iaccarino, and
  Xiao]{duraisamy2019turbulence}
Duraisamy, K., Iaccarino, G., and Xiao, H.
\newblock Turbulence modeling in the age of data.
\newblock \emph{Annual Review of Fluid Mechanics}, 51:\penalty0 357--377, 2019.

\bibitem[Emory et~al.(2013)Emory, Larsson, and Iaccarino]{emory2013modeling}
Emory, M., Larsson, J., and Iaccarino, G.
\newblock Modeling of structural uncertainties in reynolds-averaged
  navier-stokes closures.
\newblock \emph{Physics of Fluids}, 25\penalty0 (11):\penalty0 110822, 2013.

\bibitem[Gorl{\'e} et~al.(2019)Gorl{\'e}, Zeoli, Emory, Larsson, and
  Iaccarino]{gorle2019epistemic}
Gorl{\'e}, C., Zeoli, S., Emory, M., Larsson, J., and Iaccarino, G.
\newblock Epistemic uncertainty quantification for reynolds-averaged
  navier-stokes modeling of separated flows over streamlined surfaces.
\newblock \emph{Physics of Fluids}, 31\penalty0 (3):\penalty0 035101, 2019.

\bibitem[Heyse et~al.(2021{\natexlab{a}})Heyse, Mishra, and
  Iaccarino]{heyse2021data}
Heyse, J.~F., Mishra, A.~A., and Iaccarino, G.
\newblock Data driven physics constrained perturbations for turbulence model
  uncertainty estimation.
\newblock In \emph{AAAI Spring Symposium: MLPS}, 2021{\natexlab{a}}.

\bibitem[Heyse et~al.(2021{\natexlab{b}})Heyse, Mishra, and
  Iaccarino]{heyse2021estimating}
Heyse, J.~F., Mishra, A.~A., and Iaccarino, G.
\newblock Estimating rans model uncertainty using machine learning.
\newblock \emph{Journal of the Global Power and Propulsion Society},
  2021\penalty0 (May):\penalty0 1--14, 2021{\natexlab{b}}.

\bibitem[Iaccarino et~al.(2017)Iaccarino, Mishra, and
  Ghili]{iaccarino2017eigenspace}
Iaccarino, G., Mishra, A.~A., and Ghili, S.
\newblock Eigenspace perturbations for uncertainty estimation of single-point
  turbulence closures.
\newblock \emph{Physical Review Fluids}, 2\penalty0 (2):\penalty0 024605, 2017.

\bibitem[Mishra \& Iaccarino(2017{\natexlab{a}})Mishra and
  Iaccarino]{mishra2017rans}
Mishra, A. and Iaccarino, G.
\newblock Rans predictions for high-speed flows using enveloping models.
\newblock \emph{arXiv preprint arXiv:1704.01699}, 2017{\natexlab{a}}.

\bibitem[Mishra \& Iaccarino(2017{\natexlab{b}})Mishra and
  Iaccarino]{mishra2017uncertainty}
Mishra, A.~A. and Iaccarino, G.
\newblock Uncertainty estimation for reynolds-averaged navier--stokes
  predictions of high-speed aircraft nozzle jets.
\newblock \emph{AIAA Journal}, 55\penalty0 (11):\penalty0 3999--4004,
  2017{\natexlab{b}}.

\bibitem[Mishra et~al.(2019{\natexlab{a}})Mishra, Duraisamy, and
  Iaccarino]{mishra2019estimating}
Mishra, A.~A., Duraisamy, K., and Iaccarino, G.
\newblock Estimating uncertainty in homogeneous turbulence evolution due to
  coarse-graining.
\newblock \emph{Physics of Fluids}, 31\penalty0 (2):\penalty0 025106,
  2019{\natexlab{a}}.

\bibitem[Mishra et~al.(2019{\natexlab{b}})Mishra, Mukhopadhaya, Iaccarino, and
  Alonso]{mishra2019uncertainty}
Mishra, A.~A., Mukhopadhaya, J., Iaccarino, G., and Alonso, J.
\newblock Uncertainty estimation module for turbulence model predictions in
  su2.
\newblock \emph{AIAA Journal}, 57\penalty0 (3):\penalty0 1066--1077,
  2019{\natexlab{b}}.

\bibitem[Mishra et~al.(2020)Mishra, Mukhopadhaya, Alonso, and
  Iaccarino]{mishra2020design}
Mishra, A.~A., Mukhopadhaya, J., Alonso, J., and Iaccarino, G.
\newblock Design exploration and optimization under uncertainty.
\newblock \emph{Physics of Fluids}, 32\penalty0 (8):\penalty0 085106, 2020.

\bibitem[Mukhopadhaya et~al.(2020)Mukhopadhaya, Whitehead, Quindlen, Alonso,
  and Cary]{mukhopadhaya2020multi}
Mukhopadhaya, J., Whitehead, B.~T., Quindlen, J.~F., Alonso, J.~J., and Cary,
  A.~W.
\newblock Multi-fidelity modeling of probabilistic aerodynamic databases for
  use in aerospace engineering.
\newblock \emph{International Journal for Uncertainty Quantification},
  10\penalty0 (5), 2020.

\bibitem[Nigam et~al.(2021)Nigam, Mohseni, Valverde, Voronin, Mukhopadhaya, and
  Alonso]{nigam2021toolset}
Nigam, N., Mohseni, S., Valverde, J., Voronin, S., Mukhopadhaya, J., and
  Alonso, J.~J.
\newblock A toolset for creation of multi-fidelity probabilistic aerodynamic
  databases.
\newblock In \emph{AIAA Scitech 2021 Forum}, pp.\  0466, 2021.

\bibitem[Righi(2023)]{righi2023uncertainties}
Righi, M.
\newblock Uncertainties quantification in the prediction of the aeroelastic
  response of the pazy wing tunnel model.
\newblock In \emph{AIAA SCITECH 2023 Forum}, pp.\  0761, 2023.

\bibitem[Voet et~al.(2021)Voet, Ahlfeld, Gaymann, Laizet, and
  Montomoli]{voet2021hybrid}
Voet, L.~J., Ahlfeld, R., Gaymann, A., Laizet, S., and Montomoli, F.
\newblock A hybrid approach combining dns and rans simulations to quantify
  uncertainties in turbulence modelling.
\newblock \emph{Applied Mathematical Modelling}, 89:\penalty0 885--906, 2021.

\bibitem[Wu et~al.(2018)Wu, Xiao, and Paterson]{wu2018physics}
Wu, J.-L., Xiao, H., and Paterson, E.
\newblock Physics-informed machine learning approach for augmenting turbulence
  models: A comprehensive framework.
\newblock \emph{Physical Review Fluids}, 3\penalty0 (7):\penalty0 074602, 2018.

\bibitem[Xiao et~al.(2016)Xiao, Wu, Wang, Sun, and Roy]{xiao2016quantifying}
Xiao, H., Wu, J.-L., Wang, J.-X., Sun, R., and Roy, C.
\newblock Quantifying and reducing model-form uncertainties in
  reynolds-averaged navier--stokes simulations: A data-driven, physics-informed
  bayesian approach.
\newblock \emph{Journal of Computational Physics}, 324:\penalty0 115--136,
  2016.

\bibitem[Zeng et~al.(2022)Zeng, Zhang, Li, Zhang, and Yan]{zeng2022adaptive}
Zeng, F., Zhang, W., Li, J., Zhang, T., and Yan, C.
\newblock Adaptive model refinement approach for bayesian uncertainty
  quantification in turbulence model.
\newblock \emph{AIAA Journal}, pp.\  1--15, 2022.

\bibitem[Zhang(2021)]{zhang2021turbulent}
Zhang, H.
\newblock \emph{Turbulent and Non-Turbulent Interfaces in Low Mach Number
  Airfoil Flows}.
\newblock PhD thesis, Queen's University (Canada), 2021.

\end{thebibliography}
\bibliographystyle{synsml2023}

%%%%%%%%%%%%%%%%%%%%%%%%%%%%%%%%%%%%%%%%%%%%%%%%%%%%%%%%%%%%%%%%%%%%%%%%%%%%%%%
%%%%%%%%%%%%%%%%%%%%%%%%%%%%%%%%%%%%%%%%%%%%%%%%%%%%%%%%%%%%%%%%%%%%%%%%%%%%%%%
% APPENDIX
%%%%%%%%%%%%%%%%%%%%%%%%%%%%%%%%%%%%%%%%%%%%%%%%%%%%%%%%%%%%%%%%%%%%%%%%%%%%%%%
%%%%%%%%%%%%%%%%%%%%%%%%%%%%%%%%%%%%%%%%%%%%%%%%%%%%%%%%%%%%%%%%%%%%%%%%%%%%%%%
\newpage
%\appendix
%\onecolumn
%\section{You \emph{can} have an appendix here.}

%You can have as much text here as you want. The main body must be at most $4$ pages long.
%For the final version, one more page can be added.
%If you want, you can use an appendix like this one, even using the one-column format.
%%%%%%%%%%%%%%%%%%%%%%%%%%%%%%%%%%%%%%%%%%%%%%%%%%%%%%%%%%%%%%%%%%%%%%%%%%%%%%%
%%%%%%%%%%%%%%%%%%%%%%%%%%%%%%%%%%%%%%%%%%%%%%%%%%%%%%%%%%%%%%%%%%%%%%%%%%%%%%%

\end{document}